\documentclass[english,final,times,3p]{IEEEtran}
\usepackage[T1]{fontenc}
\usepackage[latin9]{inputenc}
\usepackage{xcolor}
\usepackage{pdfcolmk}
\usepackage{array}
\usepackage{float}
\usepackage{tipa}
\usepackage{amstext}
\usepackage{graphicx}
\PassOptionsToPackage{normalem}{ulem}
\usepackage{ulem}

\makeatletter

\providecommand{\tabularnewline}{\\}
\floatstyle{ruled}
\newfloat{algorithm}{tbp}{loa}
\providecommand{\algorithmname}{Algorithm}
\floatname{algorithm}{\protect\algorithmname}
\providecolor{lyxadded}{rgb}{0,0,1}
\providecolor{lyxdeleted}{rgb}{1,0,0}

\usepackage[noend]{algpseudocode}

\usepackage{tabularx}
\newcolumntype{Y}{>{\centering\arraybackslash}X}
\newcolumntype{Z}{>{\arraybackslash}X}

\@ifundefined{showcaptionsetup}{}{%
 \PassOptionsToPackage{caption=false}{subfig}}
\usepackage{subfig}
\makeatother

\usepackage{babel}
\begin{document}

\title{Computing backup forwarding rules in Software-Defined Networks}

\author{Niels L. M. van Adrichem, Farabi Iqbal and Fernando A. Kuipers\\
Network Architectures and Services, Delft University of Technology\\
Mekelweg 4, 2628 CD Delft, the Netherlands\\
\{N.L.M.vanAdrichem, M.A.F.Iqbal, F.A.Kuipers\}@tudelft.nl}
\maketitle
\begin{abstract}
The past century of telecommunications has shown that failures in
networks are prevalent. Although much has been done to prevent failures,
network nodes and links are bound to fail eventually. Failure recovery
processes are therefore needed. Failure recovery is mainly influenced
by (1) detection of the failure, and (2) circumvention of the detected
failure. However, especially in SDNs where controllers recompute network
state reactively, this leads to high delays. Hence, next to primary
rules, backup rules should be installed in the switches to quickly
detour traffic once a failure occurs. In this work, we propose algorithms
for computing an all-to-all primary and backup network forwarding
configuration that is capable of circumventing link and node failures.
Omitting the high delay invoked by controller recomputation through
preconfiguration, our proposal's recovery delay is close to the detection
time which is significantly below the 50 ms rule of thumb. After initial
recovery, we recompute network configuration to guarantee protection
from future failures. Our algorithms use packet-labeling to guarantee
correct and shortest detour forwarding. The algorithms and labeling
technique allow packets to return to the primary path and are able
to discriminate between link and node failures. The computational
complexity of our solution is comparable to that of all-to-all-shortest
paths computations. Our experimental evaluation on both real and generated
networks shows that network configuration complexity highly decreases
compared to classic disjoint paths computations. Finally, we provide
a proof-of-concept OpenFlow controller in which our proposed configuration
is implemented, demonstrating that it readily can be applied in production
networks.
\end{abstract}

\section*{}

\section{Introduction}

Modern telecommunication networks deliver a multitude of high-speed
communication services through large-scale connection-oriented and
packet-switched networks running on top of optical networks, Digital
Subscriber Lines (DSLs), cable connections or even wireless terrestrial
and satellite links. As society heavily depends on modern telecommunication
networks, much has been done to prevent network failure, e.g., by
improving the equipment environment and physical aspects of the material.
However, the past century of telecommunications has shown that network
components still fail regularly \cite{Doerr14}. Regardless of the
preventive protection measures taken, network nodes and links will
eventually malfunction and cease to function. 

In connection-oriented networks, e.g. wavelength-routed networks,
network service interruptions due to the failure of network nodes
or links can often be prevented by assigning at least two disjoint
paths from the source node to the destination node of each network
connection \cite{Kuipers12}. Connection status is then monitored
from the source node to the destination node. When the primary path
of a network connection fails, the connection can be reconfigured
to use its backup path instead. Traffic can also be sent on the primary
and backup paths of a connection concurrently, such that reconfiguration
upon the failure of the primary path is not needed. Although finding
a pair of (min-sum) disjoint paths from a source node to a destination
node is polynomially solvable \cite{suurballe1974disjoint,Suurballe84},
the returned paths may each be substantially longer than the shortest
possible path between the nodes due to the existence of trap topologies
\cite{Dunn94}. An alternative would be to find a pair of min-min
disjoint paths, where the weight of the primary path is to be minimized,
instead of the sum of the weights of both paths (min-sum). However,
the problem will then be NP-hard \cite{Guo13}.

Packet-switched networks, e.g., Ethernet or IP networks, have no connection
status since packets are forwarded in a hop-by-hop manner through
local inspection of headers at each router it traverses. Though using
disjoint paths is possible in packet-switched networks through end-to-end
liveliness detection monitoring schemes (such as Bidirectional Forwarding
Detection (BFD) \cite{rfc5880}, Ethernet OAM/CFM \cite{802.1ag-2007}
or IP Fast Reroute \cite{rfc5714}), the approach is more constrained
than in connection-oriented networks. 

In packet-switched networks, traffic can be rerouted along the primary
path, which is not possible in connection-oriented networks. Each
intermediate node along the primary path has the capability of forwarding
packets through another link interface when necessary. Furthermore,
after packets have been rerouted past the failure, packets are directed
to the shortest remaining path towards the destination, possibly by
following the remainder of the (initial) primary path that is unaffected
by the failure. However, configuring such an all-to-all configuration
requires complex forwarding rule constructions, making manual configuration
infeasible. Granular insight into the network topology is necessary,
making it difficult for traditional distributed routing protocols
to derive a correct steady state. A Software-Defined Networking (SDN)
approach may facilitate implementing such network functionality.

\begin{figure*}
\subfloat[Failure-disjoint path from $H_{1}$ to $H{}_{2}$]{\includegraphics[width=0.475\textwidth]{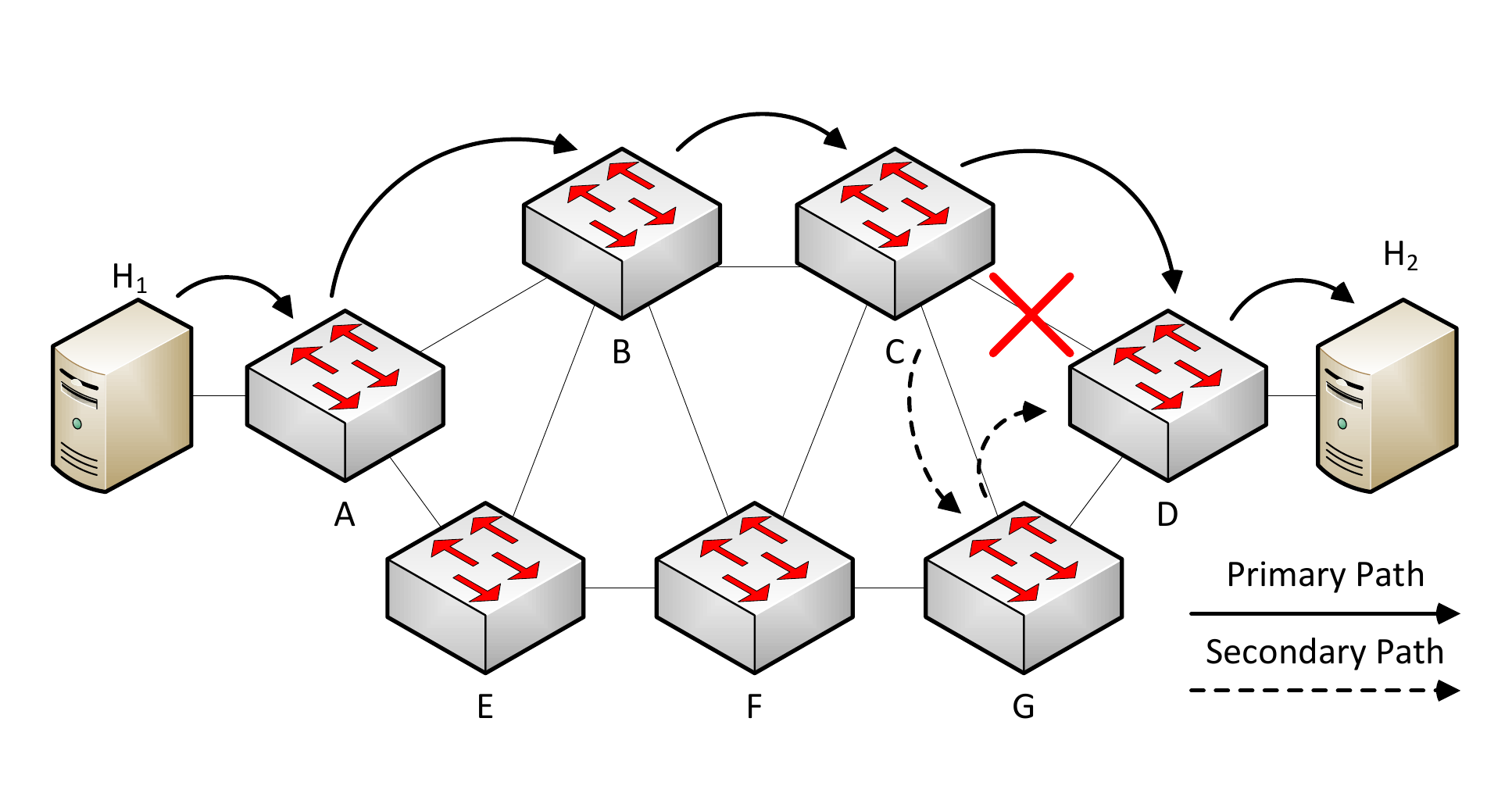}

\label{fig:FailureDisjoint}}\hfill{}\subfloat[Labels used in forwarding]{\includegraphics[width=0.475\textwidth]{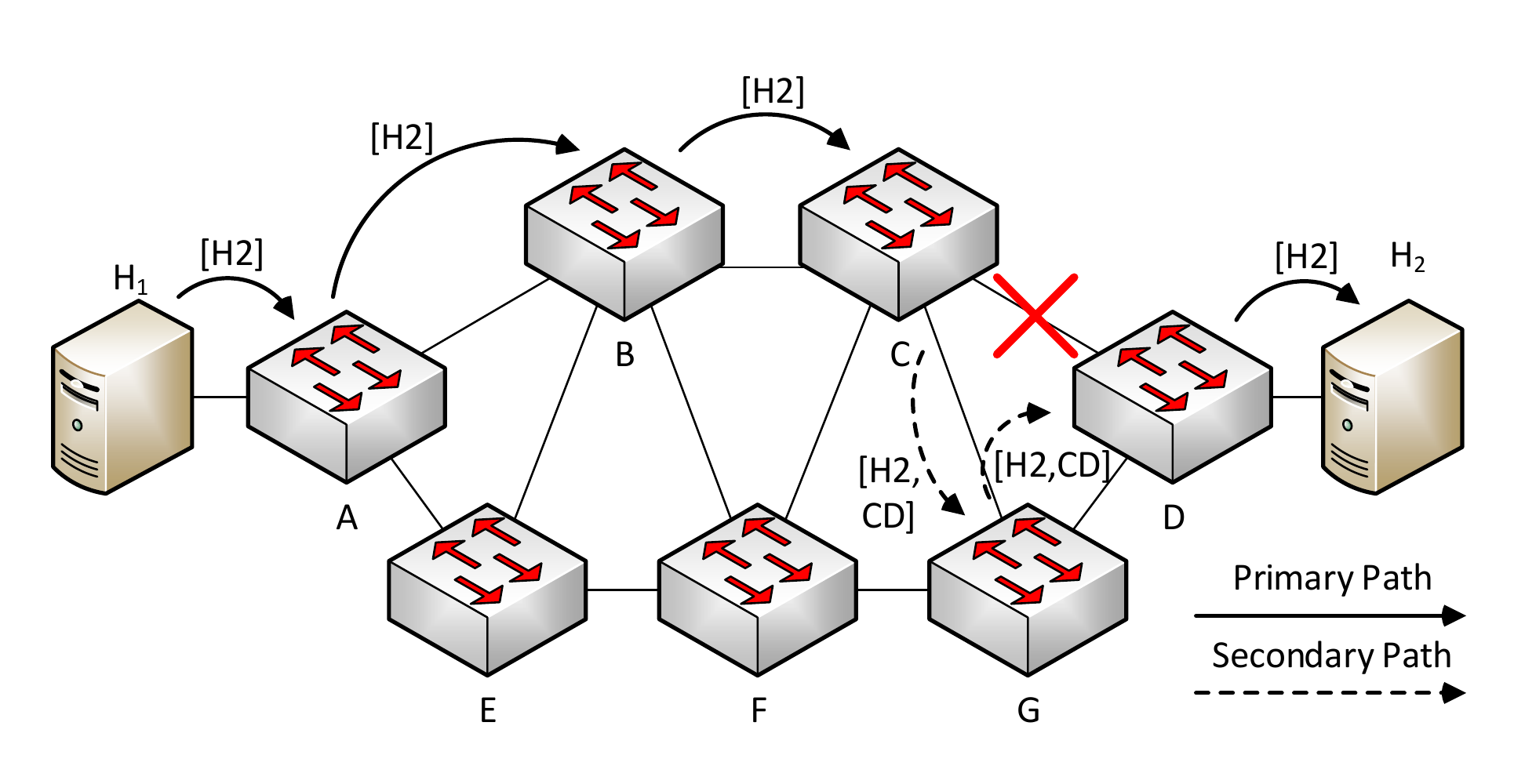}

\label{fig:FailureDisjointLabels}}

\caption{Failure disjoint paths and labels used in forwarding}
\end{figure*}

SDN enables the use of a controller for recomputing the network state
reactively upon a failure, but incurs high processing delays \cite{pdemeesterOpenFlowRecovery}.
In \cite{openflowrecovery}, we have shown that failure recovery in
OpenFlow-based SDN networks is best handled in three steps, being
1) fast failure detection through liveliness monitoring protocols,
2) failure protection through computation and configuration of backup
rules prior to failure, which is the fastest recovery approach possible
but may not deliver optimal network configuration, and 3) recomputation
of optimal network state and new backup paths as soon as the failure
detection has propagated to the network controller. Our proposal \cite{openflowrecovery}
showed very fast results, but assumed the configuration of backup
rules to be present. 

In this paper, we explore existing algorithmic solutions and propose
new ones to compute a network configuration that guarantees all-to-all
network connectivity against any single node or link failure. Our
aim is to be able to automatically configure and reconfigure any SDN
networks with failure protection schemes without human intervention. 

Our contributions in this paper are three-fold:
\begin{enumerate}
\item We derive the hard and soft constraints that should be incorporated
by a resilient routing configuration.
\item We present and evaluate algorithms for computing paths that meet those
constraints in circumventing failures.
\item We implement and experiment with the presented algorithms in an SDN
controller.
\end{enumerate}
The remainder of the paper is organized as follows. In section \ref{sec:ProblemStatement},
we formally derive a problem statement and give examples of what we
need to compute and how traditional disjoint paths algorithms fail
in doing so. Section \ref{sec:PerFailurePrecomputation} presents
our algorithmic solution for finding failure-disjoint paths, which
we evaluate and analyze in section \ref{sec:Evaluation}. Our prototype
SDN controller implementation is presented in section \ref{sec:Software}.
Section \ref{sec:RelatedWork} presents related work on finding disjoint
paths and computes their overall complexity when applied to our problem.
Finally, section \ref{sec:Conclusion} concludes the paper.

\section{Problem Statement}

\label{sec:ProblemStatement}

Figure \ref{fig:FailureDisjoint} shows an example of a shortest path
through a sample network, and a link failure between nodes C and D.
Although we are looking for an all-to-all solution, for illustration
purposes we will use the example of traffic flowing from node $H_{1}$
to node $H{}_{2}$ in the network. The primary path of the traffic,
which is the shortest path, breaks by the failure of link $l_{CD}$,
an event only noticeable by node $C$, which is an intermediate node
along the primary path. In order for the traffic to arrive at node
$H_{2},$ there must be an alternative rule to revert to at node $C$
that will ultimately route the traffic to node $H_{2}.$ In essence,
we are looking for an all-to-all solution in which all nodes are preconfigured
with backup forwarding rules to overcome any such single link or node
failure in the network. Moreover, since those rules will be computed
for each possible specific single link/node failure, both the primary
and backup paths will be as short as possible in length, which is
a big gain over standard path disjoint protection schemes. The problem
can be formally defined as follows.

\textit{Single Failure Avoidance Rule Assignment (SFARA) problem}:
Given a directed network $G$ of a set $N$ of $|N|$ nodes and a
set $L$ of $|L|$ directed links. Each link $l_{uv}\in L$ connects
nodes $u$ and $v,$ and is characterized by a link weight $\ell_{uv}$
and a boolean link status $s_{uv}$ indicating link functionality.
$s_{uv}=up$ implies that link $l_{uv}$ is functioning normally,
while $s_{uv}\neq up$ implies that link $l_{uv}$ is not functioning.
Find an overall set of primary and backup forwarding rules such that
any possible source node $x\in N$ can send packets to any possible
destination node $y\in N$ when all links are operational ($\forall l_{uv}\in L:s_{uv}=up$
), or under a single link (or node) failure ($\text{\ensuremath{\exists}!}l_{uv}\in L:s_{uv}\neq up$).

The following constraints exist for the SFARA problem: 
\begin{enumerate}
\item The status $s_{uv}$ of each link $l_{uv}\in L$ is only available
from its adjacent nodes $u$ and $v$, and may be used in the forwarding
logic of nodes $u$ and $v$. For example, $(s_{uv}=up)?(output(l_{uv})):(output(l_{uw}))$
describes the forwarding logic where node $u$ forwards packets to
node $v$ when link $l_{uv}$ is operational, or to node $w$ over
link $l_{uw}$ otherwise. Node $u$ thereby relies on node $w$ to
have a suitable backup path towards the destination.
\item A set of forwarding actions can be performed on a packet at each node,
including (a) dropping it, (b) rewriting, adding or removing any of
its labels and (c) forwarding it to the next node by outputting it
to a specific output port or link. 
\item The appropriate forwarding actions for each packet are selected from
a forwarding table based on properties such as: (a) the packet's incoming
port, (b) (wildcard) matching on packet labels such as its Ethernet
addresses, IP addresses, TCP or UDP source and destination address,
VLAN tags, MPLS labels, etc., and (c) status of the outgoing links
of the router or switch.
\end{enumerate}

\section{Per-failure precomputation for affected shortest paths}

\label{sec:PerFailurePrecomputation}

As shown earlier in figure \ref{fig:FailureDisjoint}, disjoint-path
based forwarding rules cannot instruct node $C$ on how to circumvent
the failed link. Node $C$ can only send the packet back to the source
node through crankback routing, which is an expensive process since
it uses twice the network resources from the source node to the failed
link plus the network resources on the disjoint-path. Instead, we
propose to use a detour around the failure as shown in figure \ref{fig:FailureDisjointLabels},
optimizing the primary path to the shortest path whenever possible.

We explain our algorithm for finding and configuring link-failure
disjoint paths using labeling techniques in subsection \ref{sub:LinkFailureDisjointApproach},
and later modify it to node-failure disjoint paths in subsection \ref{sub:NodeFailureDisjointApproach}.
Knowing whether a link or node failure has manifested can be difficult
since each node can only determine that an adjacent link is broken,
while the failure may only be limited to the reported link or may
include the adjacent node (and all of its links). A conservative approach
would be to assume that all link-failures imply node failures as well,
but this leads to higher detours and possibly false-negatives in determining
whether there exists a detour path to the destination node. Subsection
\ref{sub:LinkThenNodeFailureDisjointApproach} thus presents our hybrid
adaptation from the link- and node-failure disjoint paths where we
use a labelling technique to ``upgrade'' a link-failure to a node-failure
only when necessary, and adapt the forwarding strategy accordingly.
Finally, section \ref{sub:RoutingTableOptimization} discusses how
we optimize routing table complexity by removing redundant rules.

\subsection{Link-failure disjoint paths}

\label{sub:LinkFailureDisjointApproach}

Algorithm \ref{alg:per-link} presents our algorithm for computing
primary and backup forwarding rules for all possible source-destination
pairs given that at most one link is broken at any time. The algorithm
computes primary and backup forwarding rules for the whole network,
such that it is resilient to any single link failure. The algorithm
first optimizes the length of the primary path, and then optimizes
the length of the detour towards the destination node for all possible
link failures.

\begin{algorithm}
\textbf{Input:} Adjacency matrix $adj=G(N,L)$\\
\textbf{Output:} Forwarding matrix $fw$ containing primary and backup rules
\begin{algorithmic}[1]
	\State{set $fw$ to all-to-all shortest paths matrix}
	\For{each node $n \in N$}
		\For{each outgoing link $l$ of $n$}
			\State{set $tAdj$ to shadow copy of $adj$}
			\State{remove link $l$ from $tAdj$}
			\State{set $\{n'\}$ from $N$ where $nextLink=l$}
			\State{compute 1-to-$\{n'\}$ shortest paths from $tAdj$:}
			\State{\	\	\	\	store all $nextLink$ as $fw[(curNode, l)][n']$}
		\EndFor
	\EndFor
\State \Return $fw$
	
\end{algorithmic}

\caption{Per-link approach}

\label{alg:per-link}
\end{algorithm}

\begin{algorithm}
\textbf{Input:} Adjacency matrix $adj=G(N,L)$\\
\textbf{Output:} Forwarding matrix $fw$ containing primary and backup rules
\begin{algorithmic}[1]
	\State{set $fw$ to all-to-all shortest paths matrix}
	\For{each node $n \in N$}
		\For{each outgoing link $l$ of $n$}
			\State{set $tAdj$ to shadow copy of $adj$}
			\State \underline{set $n^{R}$ to node opposite of link $l$}
			\State \underline{remove node $n^{R}$ and adjacent links from $tAdj$}
			\State{set $\{n'\}$ from $N$ where $next-link=l$}
			\State{compute 1-to-$\{n'\}$ shortest paths from $tAdj$:}
			\State{\	\	\	\	store all $nextLink$ as $fw[(curNode, $\underline{$n^{R}$}$)][n']$}
		\EndFor
	\EndFor
\State \Return $fw$
\end{algorithmic}

\caption{Per-node approach, changes compared to algorithm \ref{alg:per-link}
are underlined}

\label{alg:per-node}
\end{algorithm}

\begin{figure*}
\subfloat[Node failure at node $C$ incorrectly handled by link-disjoint backup
path]{\includegraphics[width=0.475\textwidth]{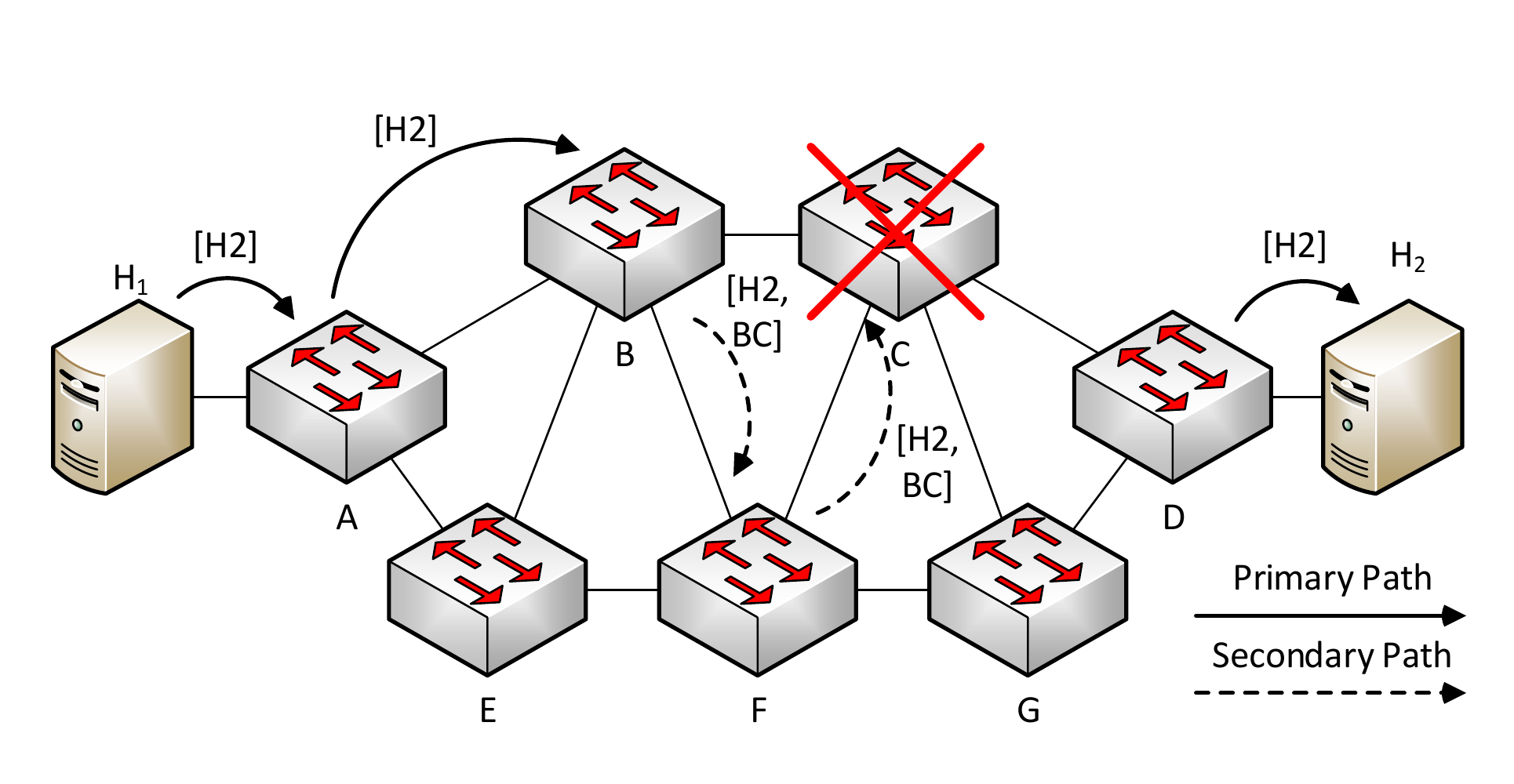}

\label{fig:NodeDisjointIncorrect}}\hfill{}\subfloat[Proper detour around failure of node $C$]{\includegraphics[width=0.475\textwidth]{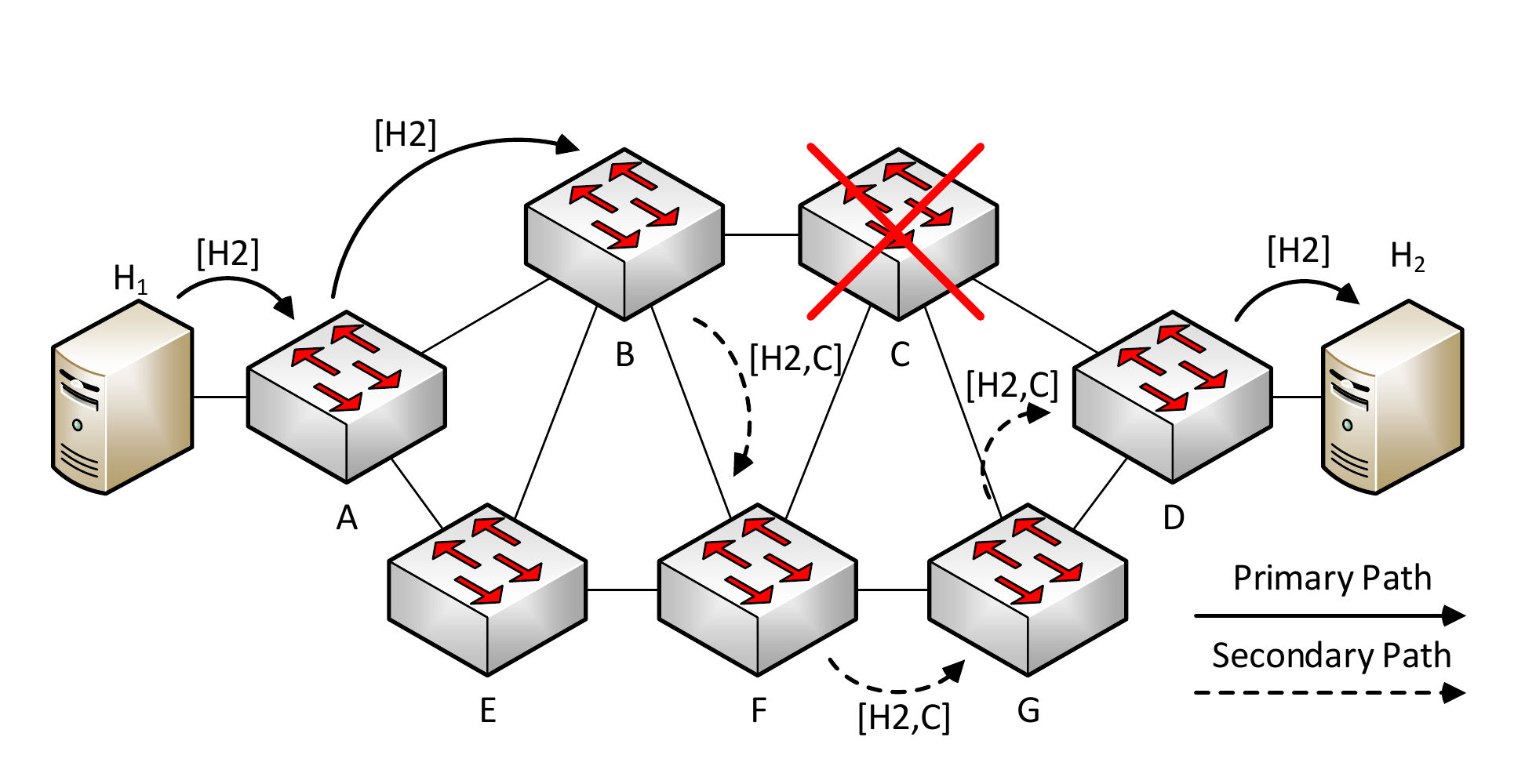}

\label{fig:NodeDisjointCorrect}}

\caption{Node failure disjointness and labels used in forwarding}
\end{figure*}

Line 1 computes a regular all-to-all shortest paths matrix, using
algorithms such as $|N|$ iterations of Dijkstra's algorithm \cite{Dijkstra59}
or the Bellman-Ford algorithm \cite{Bellman58,Ford62}, as long as
it supports the link weights in consideration\footnote{In dense graphs one may consider to use Floyd-Warshall's algorithm
\cite{Floyd62,Warshall62}. Although it is computationally more complex,
$O(|N|^{3})$, its memory complexity during computation is limited
to the size of the input adjacency matrix and output forwarding matrix
($O(|N|^{2})$).}. Lines 2 and 3 iterate through all nodes' outgoing links. Since any
link connects exactly two nodes this results in a combined complexity
of $2|L|$, leading to an intermediate complexity determined by $|N|$
times the one-to-all shortest paths computations and $\mathcal{O}(|L|)$
for the following procedure. Line 4 creates a shadow copy of the adjacency
matrix, which stores only the changes from the original. Calls to
the shadow copy check for changes first and if absent return the original
result from the original table. Since we remove at most one link,
our shadow copy suffices to be a function call to the original table
that filters out the one link before looking up the value in the matrix.
Hence, creation and lookup both have a constant complexity. Line 5
removes the link under evaluation from the shadow copy which has a
complexity of $\mathcal{O}(1)$. Line 6 selects all destinations whose
shortest paths go through the removed link. Sets containing the shortest
path destinations denoted per link can be created within a time complexity
contained by any of the suggested shortest path algorithms and hence
does not add to the overall complexity of the algorithm. Selection
of the sets is done in constant time. Finally, lines 7 and 8 compute
and store the backup paths using a regular one-to-all shortest paths
computation (such as the Dijkstra's or the Bellman-Ford algorithms)
with a slight change to the stop-criterion. First, line 7 indicates
that the algorithms may stop when all currently unreachable nodes
$\{n'\}$ have been found again, there is no need to find the shortest
paths to all nodes. Line 8 adds the found forwarding rules to the
original forwarding matrix. A distinction between the original and
backup shortest path forwarding rules from a node $n$ to its destination
forwarding rules is made by saving it under a label identifying the
specific failure, in this case link $l$. As presented in figure \ref{fig:FailureDisjointLabels},
the node that initiates sending packets through backup paths should
add a label identifying the failure it is detouring from. From this
label nodes along the backup path derive that these packets need special
treatment until they reach their destination or a shortest path that
is not affected by the failure anymore. In the latter case, the label
may be removed.

The overall complexity of the algorithm is mostly defined by the chosen
shortest path algorithm. In general, our algorithm has a worst-case
complexity of $\mathcal{O}(|N|+|L|)$ times the complexity of the
implemented shortest path algorithm, since we need $|N|$ iterations
to derive the all-to-all forwarding table and need to recompute broken
shortest paths twice for all $|L|$ links. Our solution optimizes
shortest and backup path length in sequential order. Hence, it does
not include Quality-of-Service constraints. Such functionality can
be implemented by computing primary paths using a multi-constrained
path algorithm (e.g. \cite{vanmieghem-qosconcepts}), and subsequently
computing the backup paths compared based on the remaining set of
resources. The implementation and evaluation of this solution, however,
is beyond the scope of this paper.

\subsection{Node-failure disjoint paths}

\label{sub:NodeFailureDisjointApproach}

In the case of a node failure, Algorithm \ref{alg:per-link} may not
work as the node opposite to the detected broken link is not excluded
from the backup path. Figure \ref{fig:NodeDisjointIncorrect} shows
how the selection of a link-disjoint path may send packets right back
towards the broken node. Even if node $F$ would select its link-disjoint
backup path towards node $H_{2}$, this path is not guaranteed to
be loop-free from a previous backup path. As suggested in figure \ref{fig:NodeDisjointCorrect},
in the case of a node failure we need a node-disjoint backup path
that eliminates the failed node instead of individual links from the
backup paths. Algorithm \ref{alg:per-node} presents our solution
that computes primary and backup forwarding rules for all-to-all paths
given that at most one node is broken. The algorithm computes primary
and backup forwarding rules resilient to any single node failure.
The algorithm is almost equal to Algorithm \ref{alg:per-link}, except
for minor changes. The biggest change is found in lines 5 and 6, where
instead of the removal of link $l$, its opposite node $n^{R}$ is
removed from the shadow copy. The stored label $n^{R}$ is used in
forwarding. The computational complexity remains unchanged.

\subsection{Hybrid approach}

\label{sub:LinkThenNodeFailureDisjointApproach}

The biggest problem with link-failure disjoint paths is that they
may show problems when the node opposite of the detected failed link
is broken. The node that detects link failure cannot determine whether
the link failure is a result of a single link failure or node failure
that affects all the failed node's links. The trivial solution to
use node-failure disjoint paths whenever possible may work, but implies
longer backup paths as one cannot return to the opposite node when
it is still functional and may break connectivity when there is no
node-disjoint path available. In practice, link failures occur more
than node-failures. Although the node asserting the backup path cannot
know whether a link or node failure is present, we prefer a link-failure
disjoint path whenever possible, and a node-disjoint path otherwise.

\begin{figure}
\centering{}\includegraphics[width=0.45\textwidth]{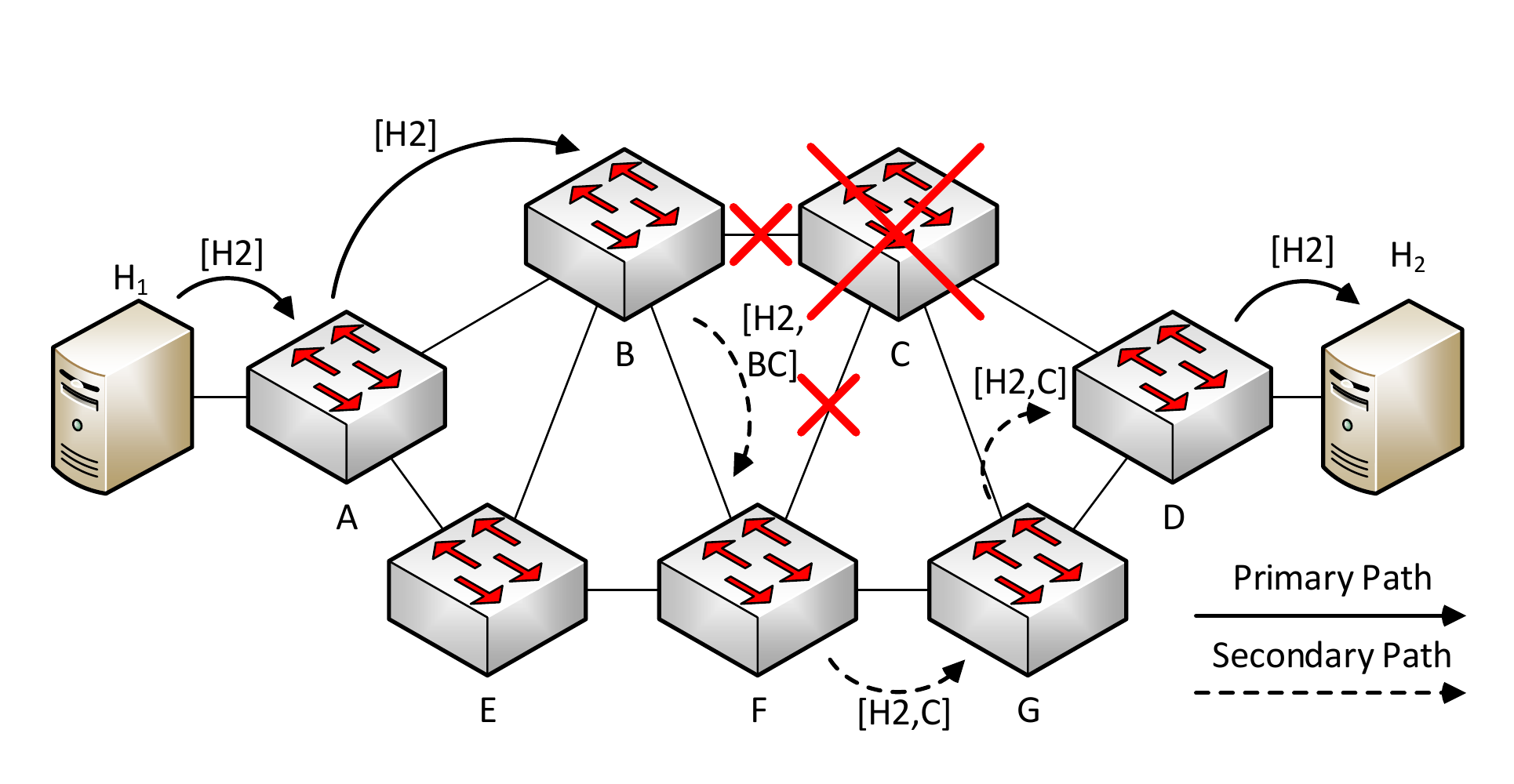}\caption{Link-, then node-failure disjoint approach and forwarding labels}
\label{fig:LinkThenNodeDisjointFigure}
\end{figure}

In order to accomplish such routing, as depicted in figure \ref{fig:LinkThenNodeDisjointFigure},
we let the asserting node assume a link-failure and act accordingly
to it by adding a label denoting link-failure and forwarding through
the link-failure disjoint path. If any node along this backup path
has a primary forwarding rule to the failed node through another of
its links, it assumes node failure based on the local link-failure
detection combined with the label on the incoming packets indicating
it is not the first broken link of that node. Furthermore, this knowledge
is added to the attached label. When every attached label of a failed
link is a concatenation of its interconnecting nodes ($\left\{ u,v\right\} $),
a forwarding rule wildcard match such as $\left\{ *,v\right\} $ can
detect previous link failures to node $v$.

To compute these rules, we compute both node- and link-failure disjoint
paths and place these using their unique labels in the shared forwarding
matrix. Note that the initial forwarding matrix only needs to be computed
once, and removal of links and nodes and their respective recomputations
may occur sequentially. This procedure runs in the same worst-case
time complexity as the previous two algorithms.

\subsection{Routing Table Optimization}

\label{sub:RoutingTableOptimization}

\begin{figure}
\centering{}\includegraphics[width=0.41\textwidth]{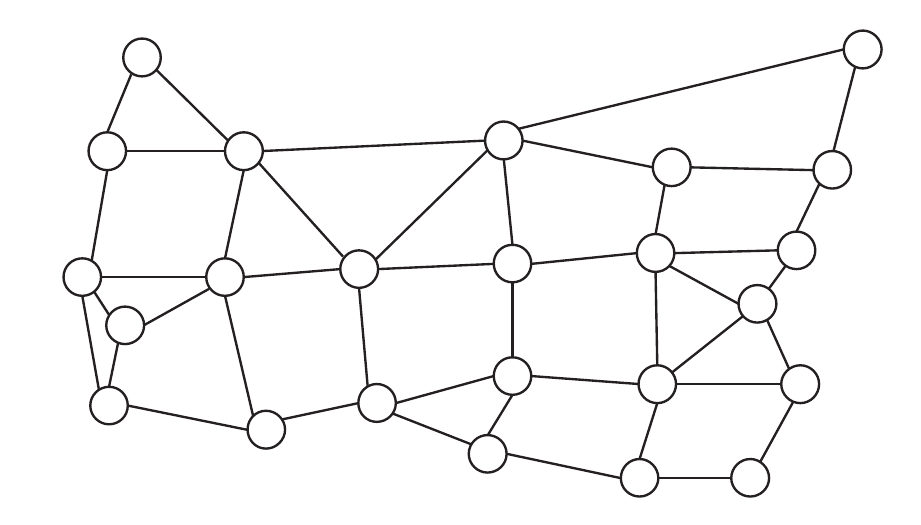}\caption{USnet topology with 24 nodes and 43 links}
\label{fig:USnet}
\end{figure}

The procedures described in the previous subsections looks for min-min
link-, node- and hybrid-failure disjoint paths. However, without optimizing
forwarding rule complexity, this results in a state explosion of forwarding
rules. While the realistic \emph{USnet} topology (shown in figure
\ref{fig:USnet}) initially has a total of $552$ forwarding rules
($23$ per switch, one for each destination), our link-, node- and
hybrid-failure disjoint approaches change most of these from regular
output actions to group tables and respectively leads to $1606$,
$2078$ and $3684$ (sum of previous two) additional entries in the
forwarding matrix $fw$. Although switches often have very large forwarding
tables for Layer 2 matching, the number of TCAM entries in a switch
for multiple field matches as performed in OpenFlow lies in the order
of $1000$ to $10000$ rules \cite{OpenFlowSwitchingPerformance}.
Hence, it is necessary to minimize the number of forwarding rules
to allow applicability in larger networks.

Considering that detoured packets at a certain point follow the default
shortest paths from intermediate nodes on the backup path to destination,
unaffected by the found failure, a first optimization is found by
removing the failure-identifying label once a suitable default shortest
path is found, leading to an addition of only $487$ and $576$ node-
or link-failure disjoint entries, which is a big improvement. Since
the hybrid-failure disjoint path may not revert to a shortest path
before a potential node-failure is omitted, we find an additional
$1293$ hybrid-failure disjoint entries, which, although larger than
the sum of the previous two, is still a factor three lower than before.

Moreover, if we consider the USnet topology to be unweighted, hence
introducing multiple shortest paths, we find an additional $621$
and $741$ node- and link-failure disjoint entries, which is larger
than its weighted counter result, indicating that it is important
for resilience in a network to have unique shortest paths.

We further optimize rulespace utilization by removing link-failure
disjoint forwarding rules in the hybrid computation when they are
equal to their respective node-failure disjoint rule, leading to a
decreased number of $847$ additional entries. A more extensive evaluation
of our proposal compared to fully disjoint paths is presented in section
\ref{sec:Evaluation}.

\section{Evaluation}

\label{sec:Evaluation}

In this section, we study the performance of our algorithms through
simulation in three network topologies, Erdös-Rényi random networks
\cite{Erdos59}, lattice networks, and Waxman networks \cite{Waxman88}.
For our generated Erdös-Rényi random networks, we choose $\frac{2log|N|}{|N|}$
as the probability for link existence, since the network will almost
surely be connected when the probability for link existence exceeds
$\frac{(1+\text{\textepsilon})log|N|}{|N|}$, where $\text{\textepsilon>0}$.
In the lattice network, all interior nodes have a degree of four and
the exterior nodes are connected to their closest exterior neighboring
nodes. The lattice network is useful in representing grid-based networks,
which may resemble the inner core of an ultra-long-reach optical data
plane system \cite{Moral01}. We choose a square lattice network of
$i\times i$ dimension, where $i=\sqrt{|N|}$, for our generated lattice
networks. The Waxman network is frequently used to model communication
networks and the Internet topology \cite{Naldi05}, due to its unique
property of decaying link existence over distance. In the Waxman network,
nodes are uniformly positioned in the plane, and link existence is
reflected by $ie^{\frac{\ell_{uv}}{ja}}$, where $\ell_{uv}$ is the
Euclidean distance between nodes $u$ and $v$, $a$ is the maximum
distance between any two nodes in the plane, and $i$ and $j$ can
vary between 0 to 1. We set $i=0.5$ and $j=0.5$ since higher $i$
leads to higher link densities, and lower $j$ leads to shorter links.
We consider only two-connected generated graphs, such that the network
can never be disconnected by a single node or link failure. In the
Erdös-Rényi and lattice networks, each link has a random link weight
between $0$ and $1$. No self-loops or parallel links are allowed.
Simulations were conducted on an Intel(R) Core i7-3770K 3.50 GHz machine
with 16GB RAM memory, and all results are averaged over a $1000$
runs and grouped by the network sizes 9, 16, 25, 36, 49, 64, 81 and
100 due to the dimension of the lattice network.

We compute and compare the results of different disjoint algorithms,
being our link-, node- and hybrid-failure disjoint approaches and
min-sum pairs of fully link- and node-disjoint paths. More specifically,
we measure:
\begin{enumerate}
\item the total number of flow entries
\item the amount of flow entries that forward to a Group table entry
\item the number of distinct Group table entries
\item the average primary path length
\item the averages of

\begin{enumerate}
\item the average, minimal and maximal backup path length for each node
pair and
\item the average, minimal and maximal crankback length for experienced
backup paths.
\end{enumerate}
\end{enumerate}
We compute link-, node- and hybrid-failure disjoint paths according
to our approach and link- and node-disjoint paths according to Bhandari's
algorithm\footnote{Note that any other min-sum disjoint paths algorithm renders the same
results, given that solutions are unique or an equal tossing method
is used.} for the generated networks, and calculate the enumerated values for
these paths.

\begin{figure}
\includegraphics[width=1\columnwidth]{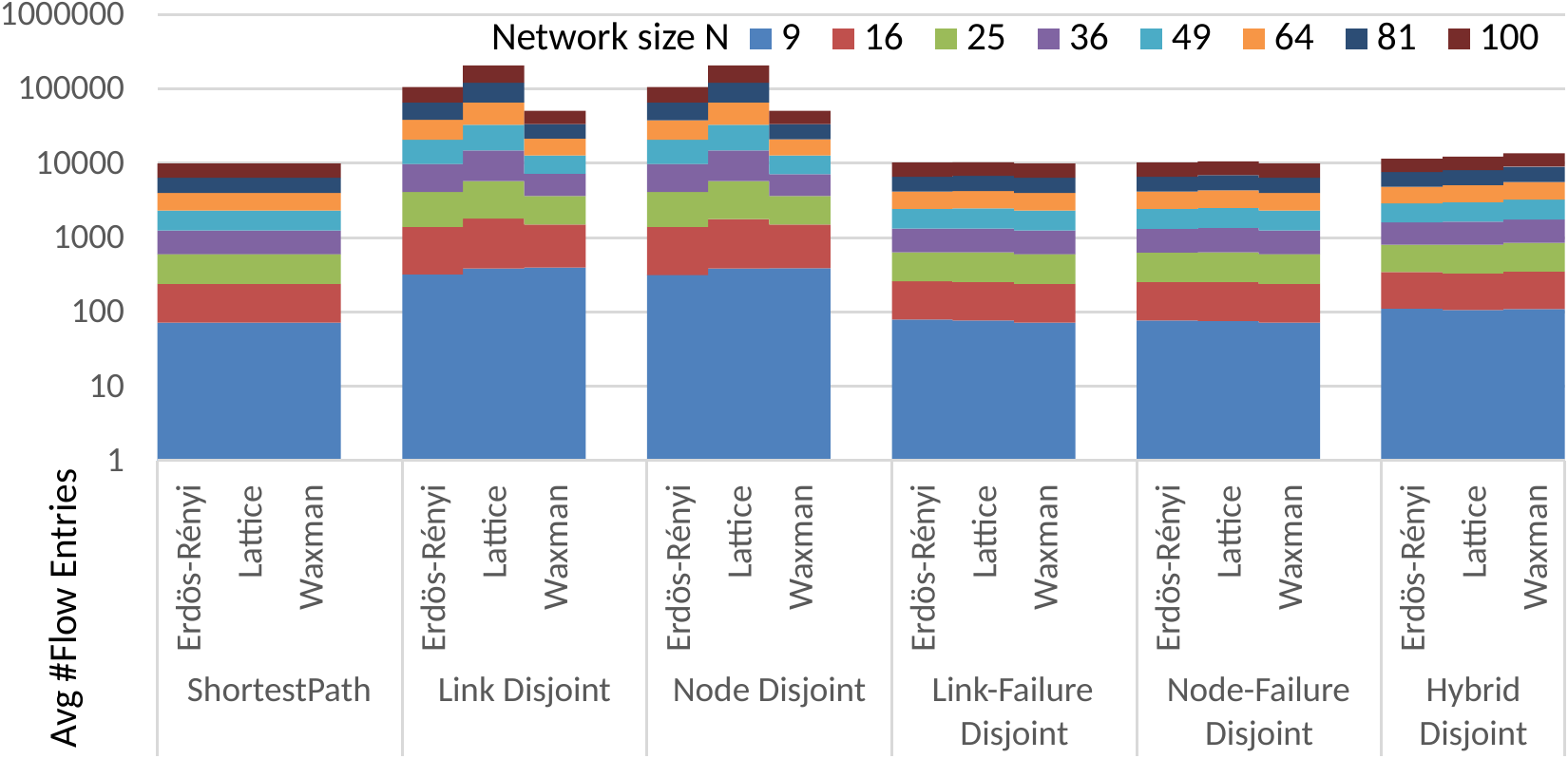}\caption{Average number of Flow entries in each network, categorized per network
type and disjoint computation and incrementally stacked per network
size}

\label{fig:FlowEntries}
\end{figure}

\begin{figure}
\includegraphics[width=1\columnwidth]{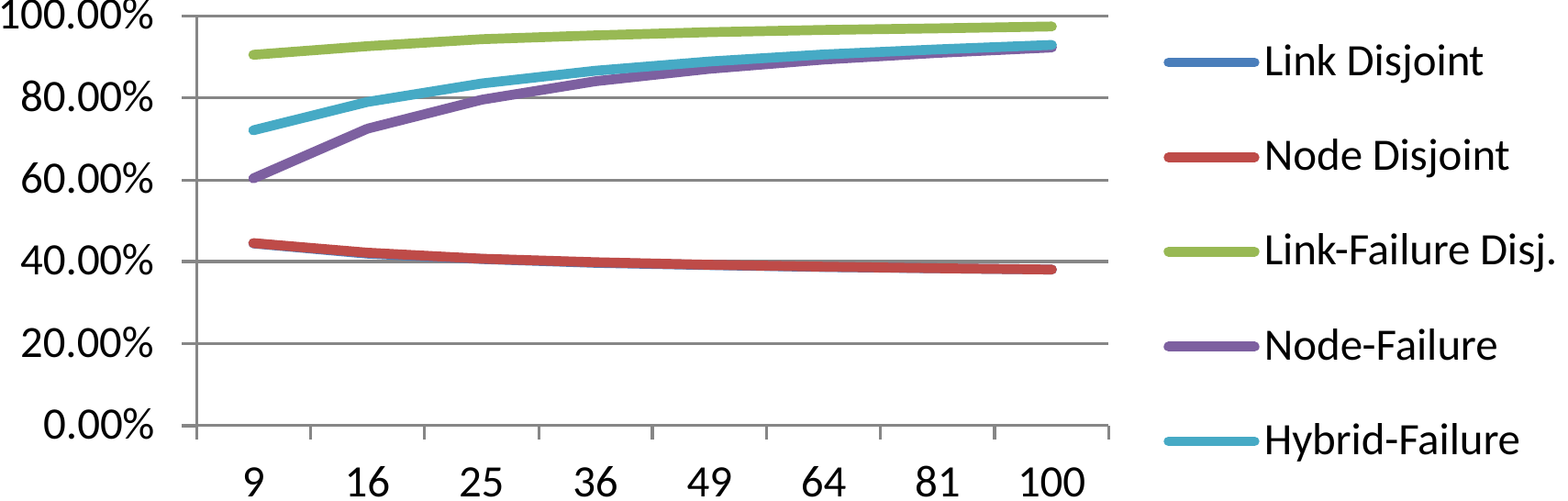}\caption{Ratio of forwards to Group table entries for Erdös-Rényi generated
random networks of size $N=100$ nodes}

\label{fig:RatioGroupEntries}
\end{figure}

\begin{table*}
\centering{}\caption{Results for the evaluated algorithms and networks of network size
$N=100$}
\begin{tabular}{|c|c|c|c|>{\centering}p{0.15\columnwidth}|>{\centering}p{0.15\columnwidth}|c|c|>{\centering}p{0.15\columnwidth}|c|}
\hline 
Network & Disjoint & Flow Entries & Distinct Groups & Primary Path Ratio & Backup Path Ratio & - Min & - Max & Crankback Ratio & - Max\tabularnewline
\hline 
\hline 
Erdös-Rényi & Link & 106852.298 & 1908.580 & 1.014 & 2.146 & 1.419 & 2.787 & 0.363 & 0.684\tabularnewline
\hline 
Erdös-Rényi & Node & 106632.844 & 1913.405 & 1.015 & 1.956 & 1.434 & 2.402 & 0.261 & 0.484\tabularnewline
\hline 
Erdös-Rényi & Link-Failure & 10160.248 & 1187.016 & 1.000 & 1.512 & 1.218 & 1.887 & 0.035 & 0.123\tabularnewline
\hline 
Erdös-Rényi & Node-Failure & 10149.929 & 1096.475 & 1.000 & 1.471 & 1.254 & 1.733 & 0.024 & 0.079\tabularnewline
\hline 
Erdös-Rényi & Hybrid-Failure & 11451.396 & 1388.225 & 1.000 & 1.547 & 1.277 & 1.914 & 0.023 & 0.076\tabularnewline
\hline 
Lattice & Link & 207585.132 & 1002.772 & 1.039 & 2.259 & 1.369 & 3.101 & 0.445 & 0.866\tabularnewline
\hline 
Lattice & Node & 207642.592 & 1006.752 & 1.050 & 2.190 & 1.403 & 2.919 & 0.394 & 0.758\tabularnewline
\hline 
Lattice & Link-Failure & 10381.567 & 571.113 & 1.000 & 1.283 & 1.095 & 1.525 & 0.016 & 0.088\tabularnewline
\hline 
Lattice & Node-Failure & 10663.192 & 542.702 & 1.000 & 1.308 & 1.127 & 1.538 & 0.012 & 0.065\tabularnewline
\hline 
Lattice & Hybrid-Failure & 12320.495 & 735.551 & 1.000 & 1.331 & 1.131 & 1.612 & 0.013 & 0.067\tabularnewline
\hline 
Waxman & Link & 50885.088 & 6208.709 & 1.000 & 1.570 & 1.049 & 2.105 & 0.260 & 0.528\tabularnewline
\hline 
Waxman & Node & 50572.606 & 6247.912 & 1.000 & 1.359 & 1.147 & 1.576 & 0.106 & 0.214\tabularnewline
\hline 
Waxman & Link-Failure & 9900 & 3874.098 & 1.000 & 1.070 & 1.032 & 1.117 & 0.000 & 0.001\tabularnewline
\hline 
Waxman & Node-Failure & 9900 & 3268.110 & 1.000 & 1.152 & 1.139 & 1.166 & 0.000 & 0.000\tabularnewline
\hline 
Waxman & Hybrid-Failure & 13671.039 & 4660.458 & 1.000 & 1.163 & 1.147 & 1.180 & 0.000 & 0.000\tabularnewline
\hline 
\end{tabular}\label{tbl:ResultsN100}
\end{table*}

\begin{figure}
\includegraphics[width=1\columnwidth]{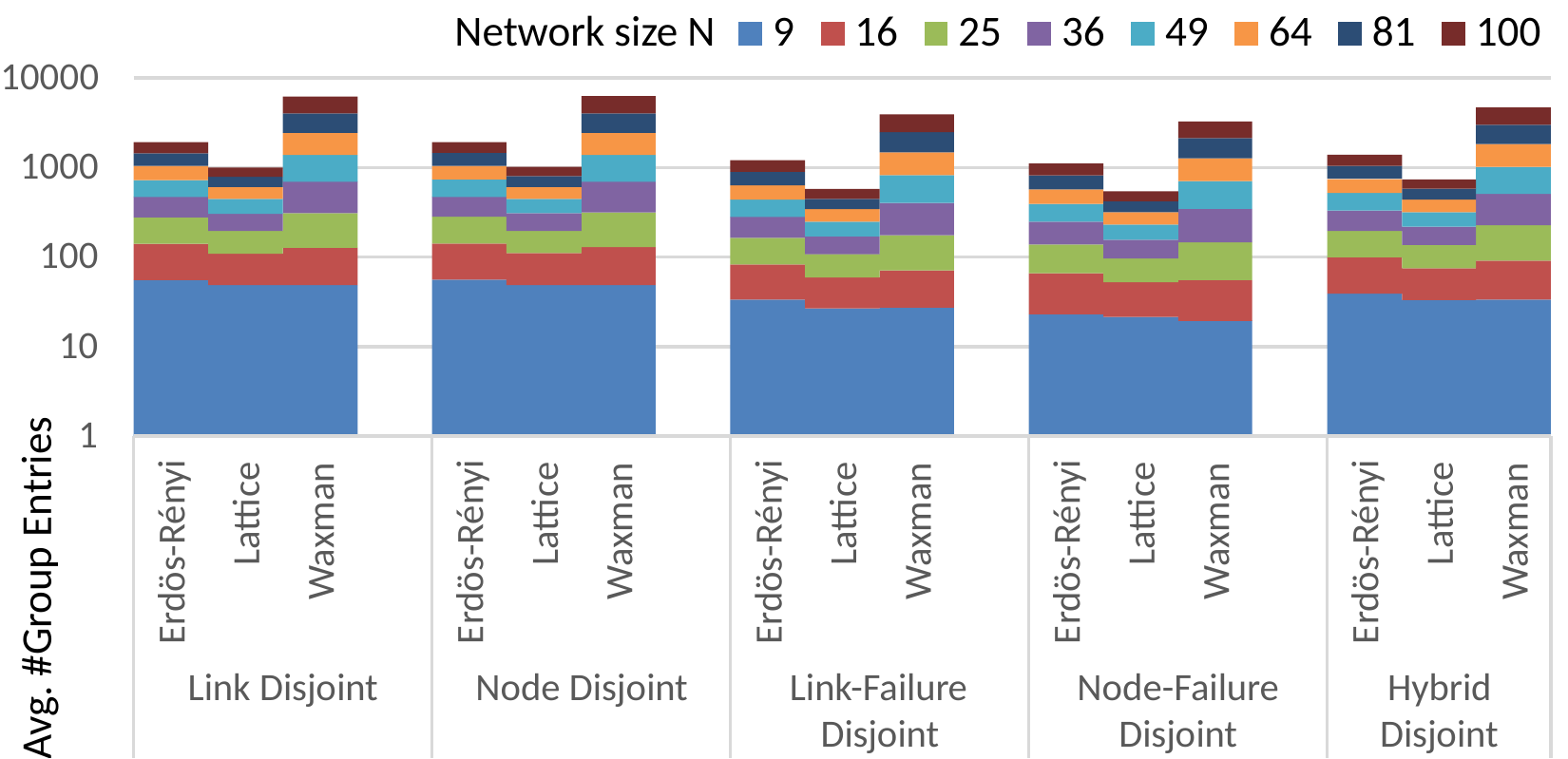}\caption{Average number of distinct Group Entries in each network, categorized
per network type and of disjoint computation and incrementally stacked
per network size}

\label{fig:DistinctGroups}
\end{figure}

Figure \ref{fig:FlowEntries} presents the average number of Flow
table entries for each generated network. A regular shortest paths
computation always generates exactly $|N|(|N|-1)$ Flow table entries
(from each node to each other node). This number increases when more
complex path computations are used. Specifically, we see a strikingly
high increase in Flow table entries when fully-disjoint paths are
used, which is caused by the fact that each forwarding rule has to
take both source and destination into account for primary path forwarding,
as well as the incoming port for crankback routing. As also shown
in table \ref{tbl:ResultsN100}, our failure-disjoint proposal shows
an increase in Flow table entries varying from $15.7\%$ to $38\%$
for a network size of $N=100$ nodes, whereas for the fully-disjoint
computations this is limited from no increase to $7.7\%$. Given that
fully-disjoint paths lead to an increased table usage by a factor
of $21$, our method appears to be much more conservative in Flow
table usage. Whereas we found that our proposal uses significantly
less flow table entries, figure \ref{fig:RatioGroupEntries} shows
up to $94\%$ of these are forwarded to Group table entries compared
to a worst case of $44\%$ for a fully disjoint path. Although this
looks like a significant increase, the absolute number of Flow entries
forwarding to Group table entries remains much lower in all cases.
Moreover, table \ref{tbl:ResultsN100} and figure \ref{fig:DistinctGroups}
show that our proposal contains a significantly lower usage of distinct
Group table entries in each network, which are considered scarce resources.

\begin{figure}
\includegraphics[width=1\columnwidth]{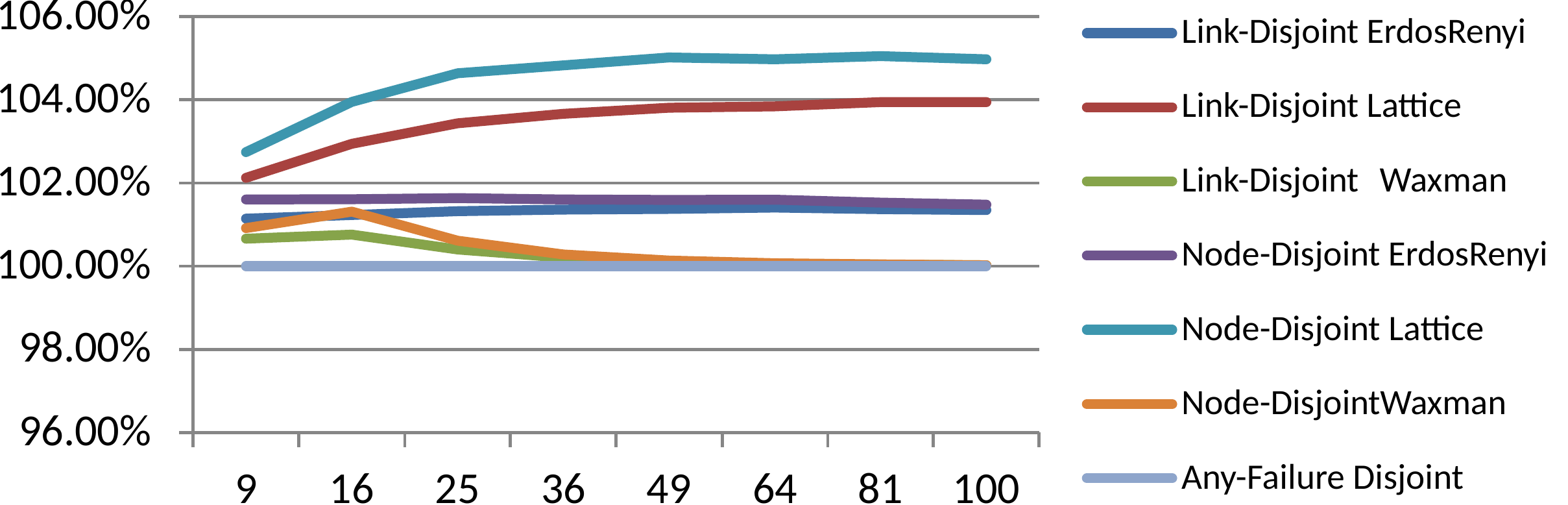}\caption{Increase in primary path per network size, categorized by algorithm
and network type.}
\label{fig:PrimaryPathIncrease}
\end{figure}

Besides a smaller configuration complexity, also the primary paths
taken are better. While the primary path in our proposal always defaults
to the shortest path, figure \ref{fig:PrimaryPathIncrease} shows
that using fully-disjoint paths leads to an increase of primary path
lengths of up to $5.0\%$, and thus incurs higher network operation
costs. Although the increase of primary path length of disjoint paths
in most cases grows and at a certain point seems to stabilize, with
Waxman generated networks the path increase decreases over time implying
that the design of the network has implications for the relative cost
of robustness.

\begin{figure}
\includegraphics[width=1\columnwidth]{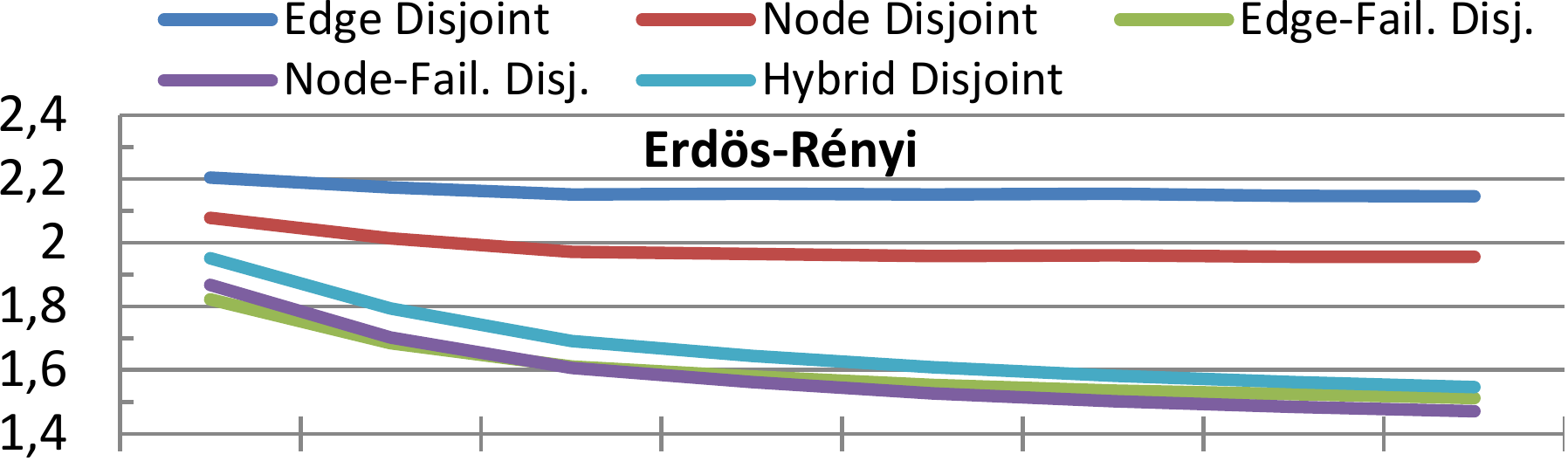}

\includegraphics[width=1\columnwidth]{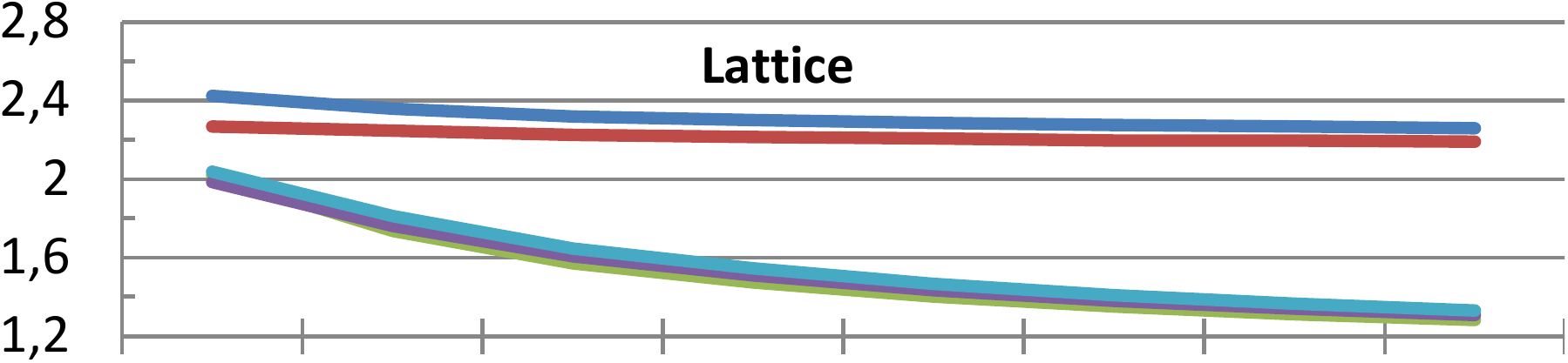}

\includegraphics[width=1\columnwidth]{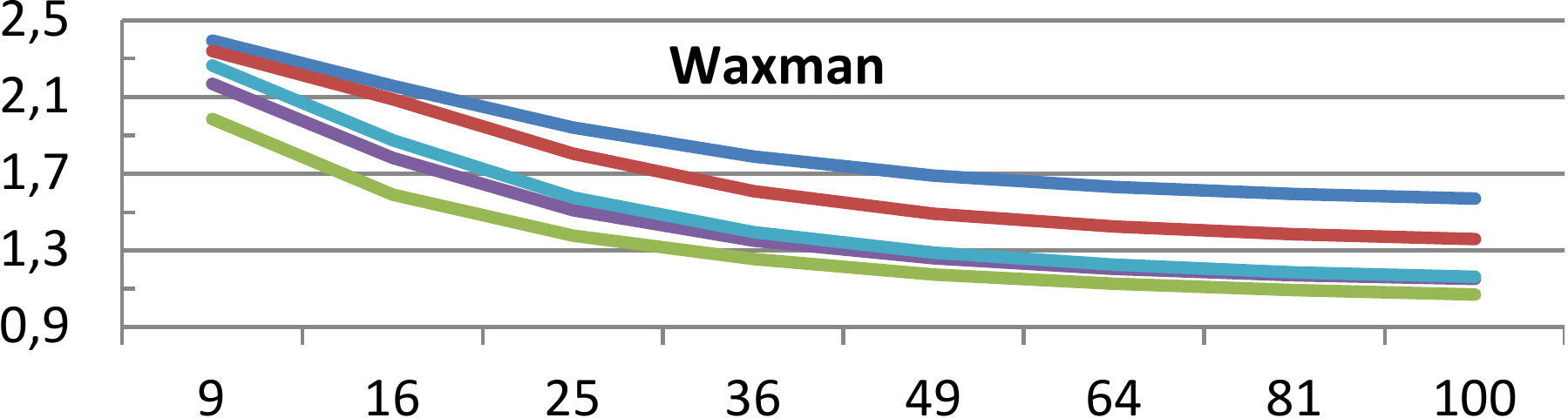}

\caption{Average length of backup paths relative to length of their respective
shortest path}
\label{fig:SecondaryPathRatio}

\includegraphics[width=1\columnwidth]{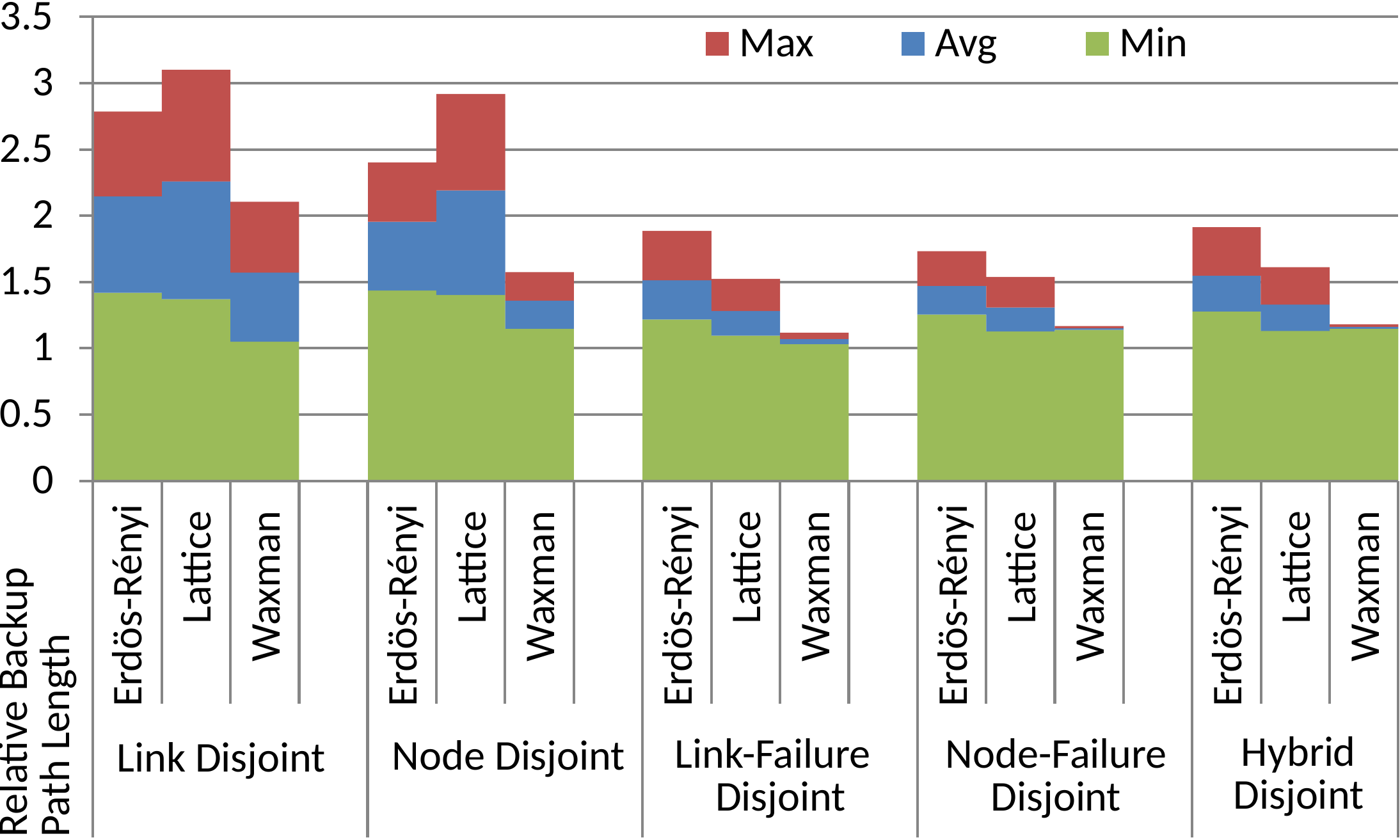}

\caption{Comparison between minimal, average and maximal relative backup path
lengths for networks of size $N=100$}
\label{fig:SecondaryPathRatioMinMax}
\end{figure}

\begin{figure}
\includegraphics[width=1\columnwidth]{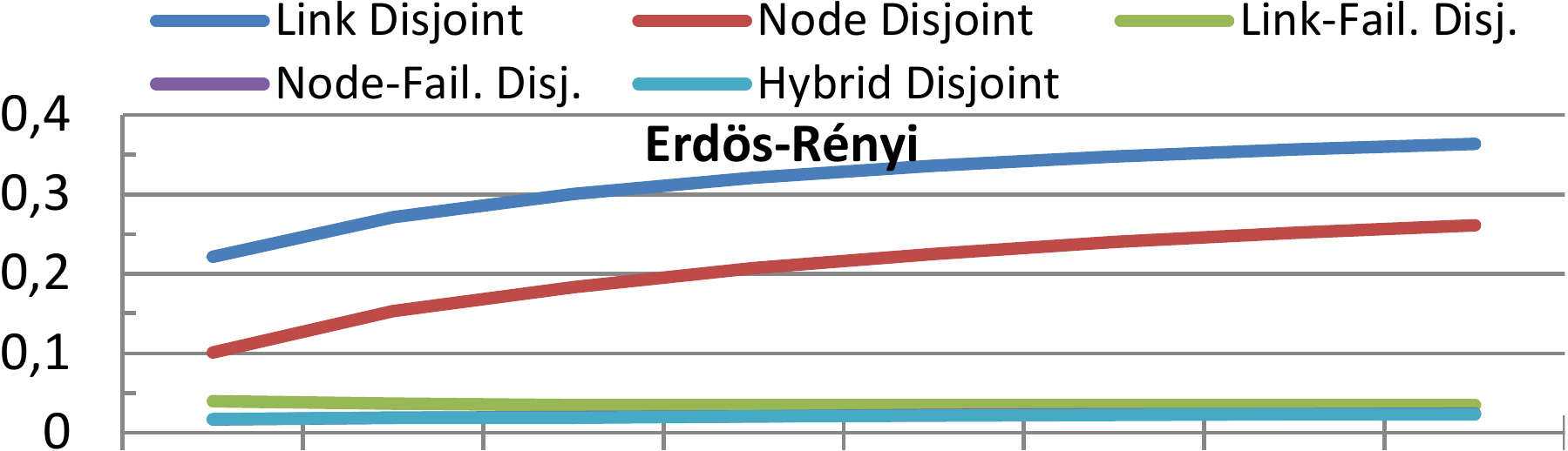}

\includegraphics[width=1\columnwidth]{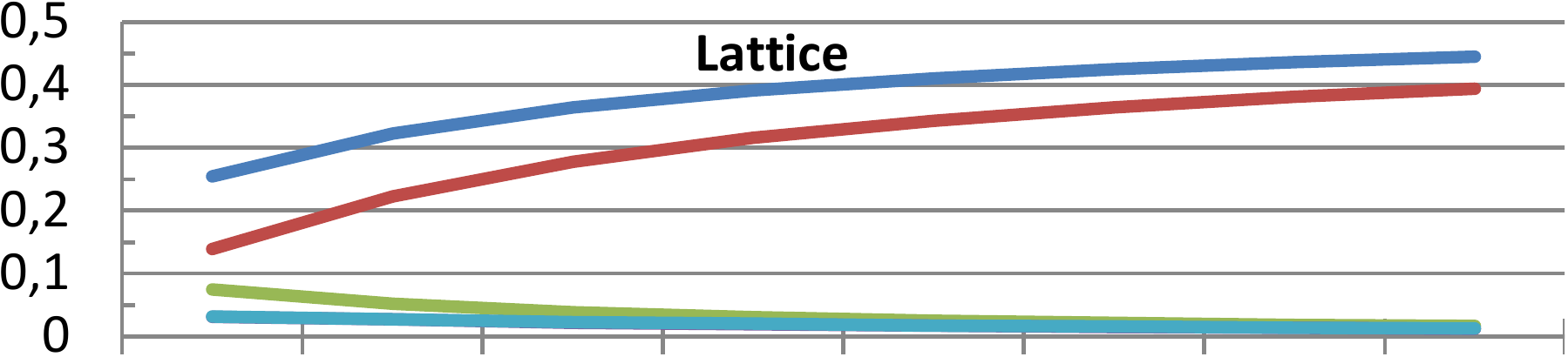}

\includegraphics[width=1\columnwidth]{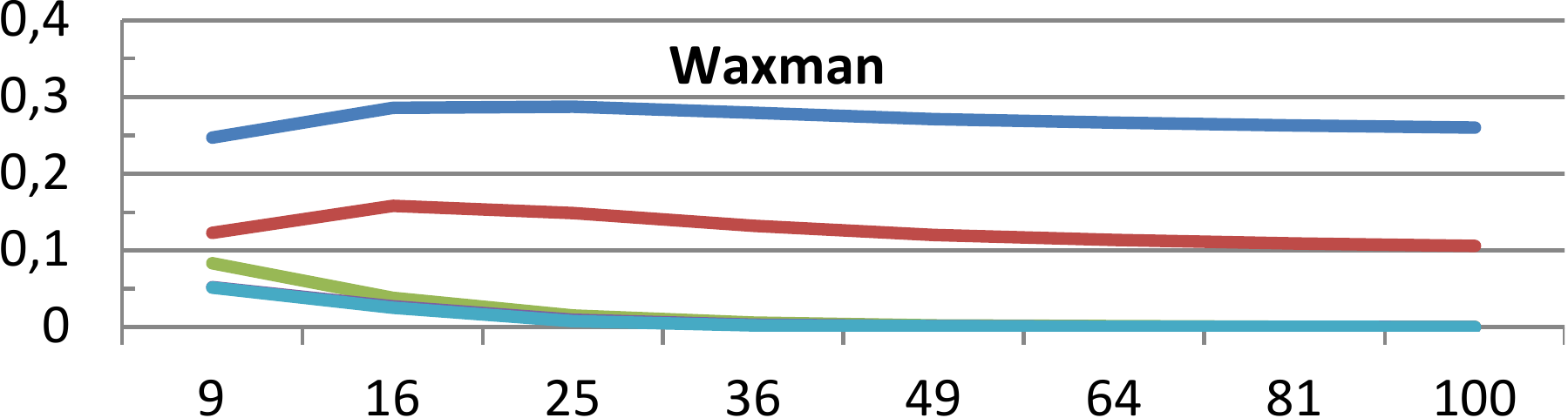}

\caption{Average length of crankback paths relative to length of their respective
shortest path}
\label{fig:CrankbackRatio}

\includegraphics[width=1\columnwidth]{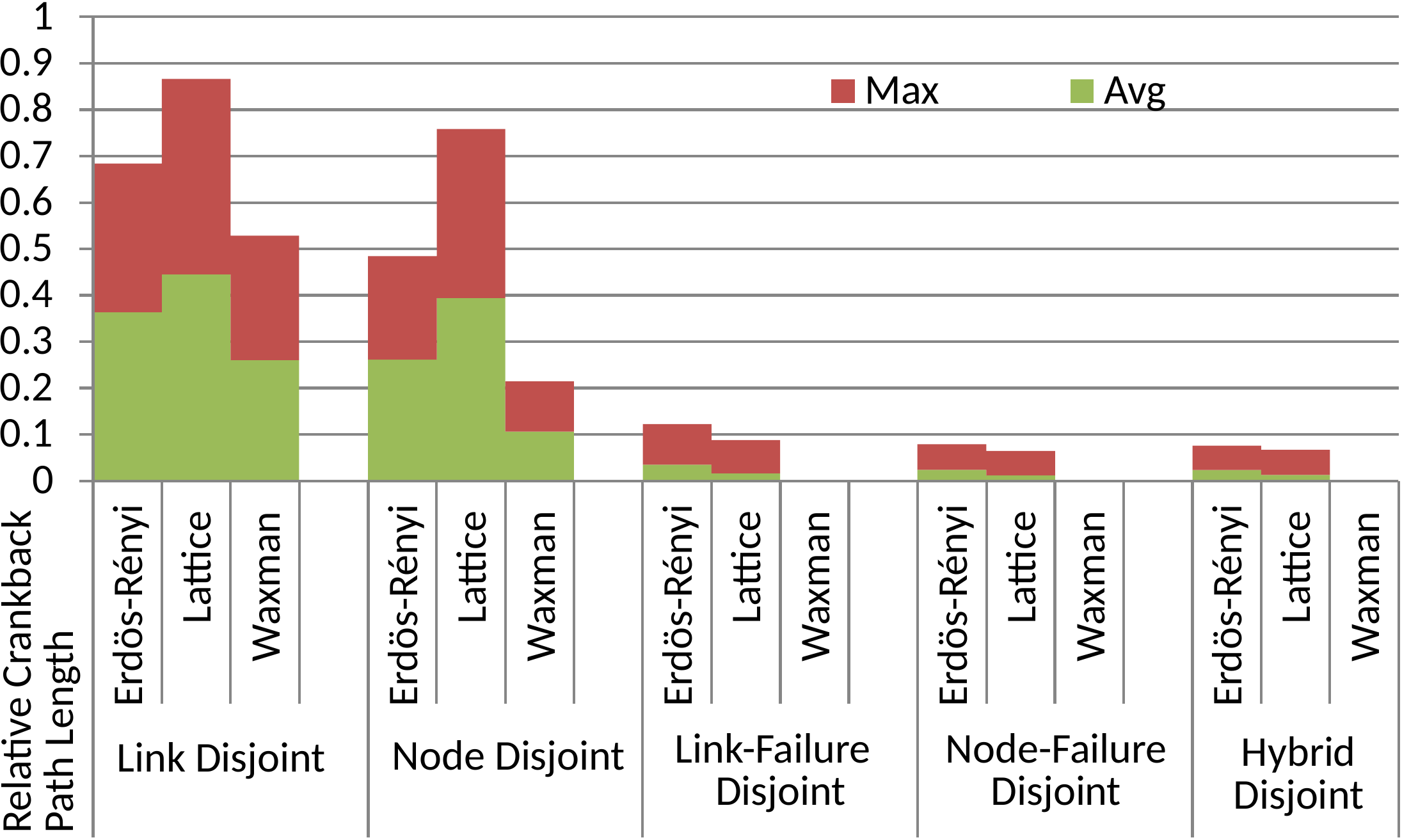}

\caption{Comparison between minimal, average and maximal relative crankback
lengths for networks of size $N=100$}
\label{fig:CrankbackRatioMinMax}
\end{figure}

Figure \ref{fig:SecondaryPathRatio} shows that besides a shorter
primary path, our proposal on average also has significantly shorter
average backup paths. In order to determine the average backup path
for a node pair, we took its primary path and for each link or node
on the path computed the length of the path if that specific link
or node would fail and averaged accordingly. Hence, as figure \ref{fig:SecondaryPathRatioMinMax}
shows, the average backup path deviates significantly based on the
link that fails. Especially the fully-disjoint paths suffer from a
high deviation due to the high order of crankback routing that is
involved when a link further down the primary path breaks. Figures
\ref{fig:CrankbackRatio} and \ref{fig:CrankbackRatioMinMax} additionally
show that the ratio and deviation of crankback paths is much larger
for fully-disjoint paths than for our approach. Furthermore, crankback
paths only exist temporarily in our proposal, since the controller
reconfigures the network by applying the protection scheme to its
newly established topology once it is notified of the failure, thereby
removing existing crankback subpaths from the shortest paths.

The hybrid-failure disjoint path lengths are only shown for a node
failure, since the path lengths for a respective link failure are
equal to the results in the link-disjoint approach by design. Although
the number of Flow and Group table entries, as well as the secondary
path and crankback length for node failures slightly increases in
the hybrid-failure approach, we claim this number is justified by
the merits of shorter paths for the more often occurring link failures.

Although no exact measurements were made, we found that our proposal
had a much faster computation time than its fully-disjoint counterpart.
Our hybrid approach in general took $4$ seconds to finish, compared
to $20$ seconds in Lattice networks and even up to a minute in Erdös-Rényi
and Waxman generated networks for the fully-disjoint approach. Hence,
our implementation is much faster in computing a new network configuration
that offers protection from a possible next failure.

\section{Software Implementation}

\label{sec:Software}

\begin{figure*}
\subfloat[The disjoint paths from $H_{1}$ to $H{}_{2}$ do not instruct node
$C$ how to handle the link failure]{\includegraphics[width=0.475\textwidth]{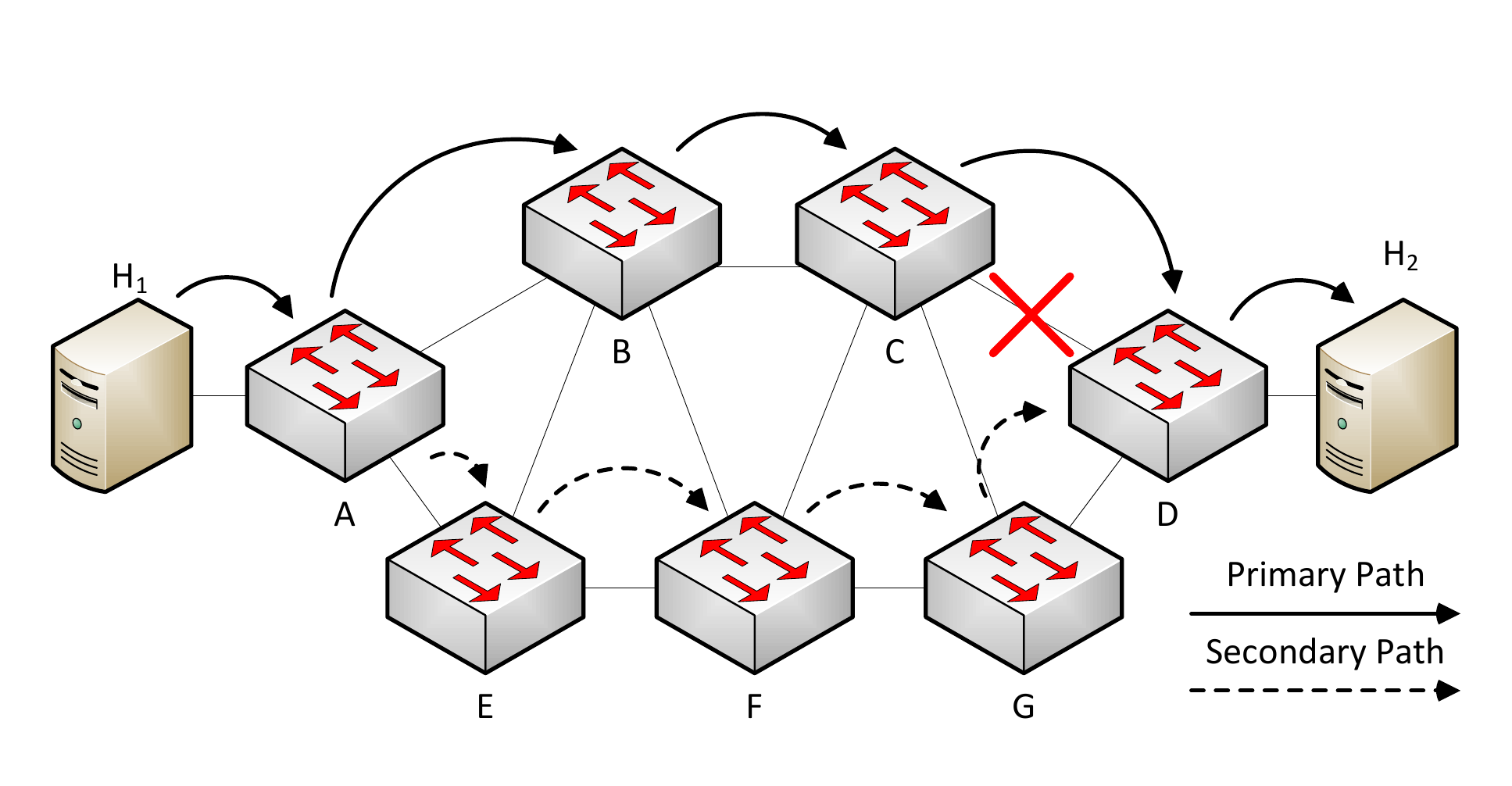}

\label{fig:DisjointPathError}}\hfill{}\subfloat[Using disjoint paths, the link failure of $l_{cd}$ can only be omitted
through crank back routing]{\includegraphics[width=0.475\textwidth]{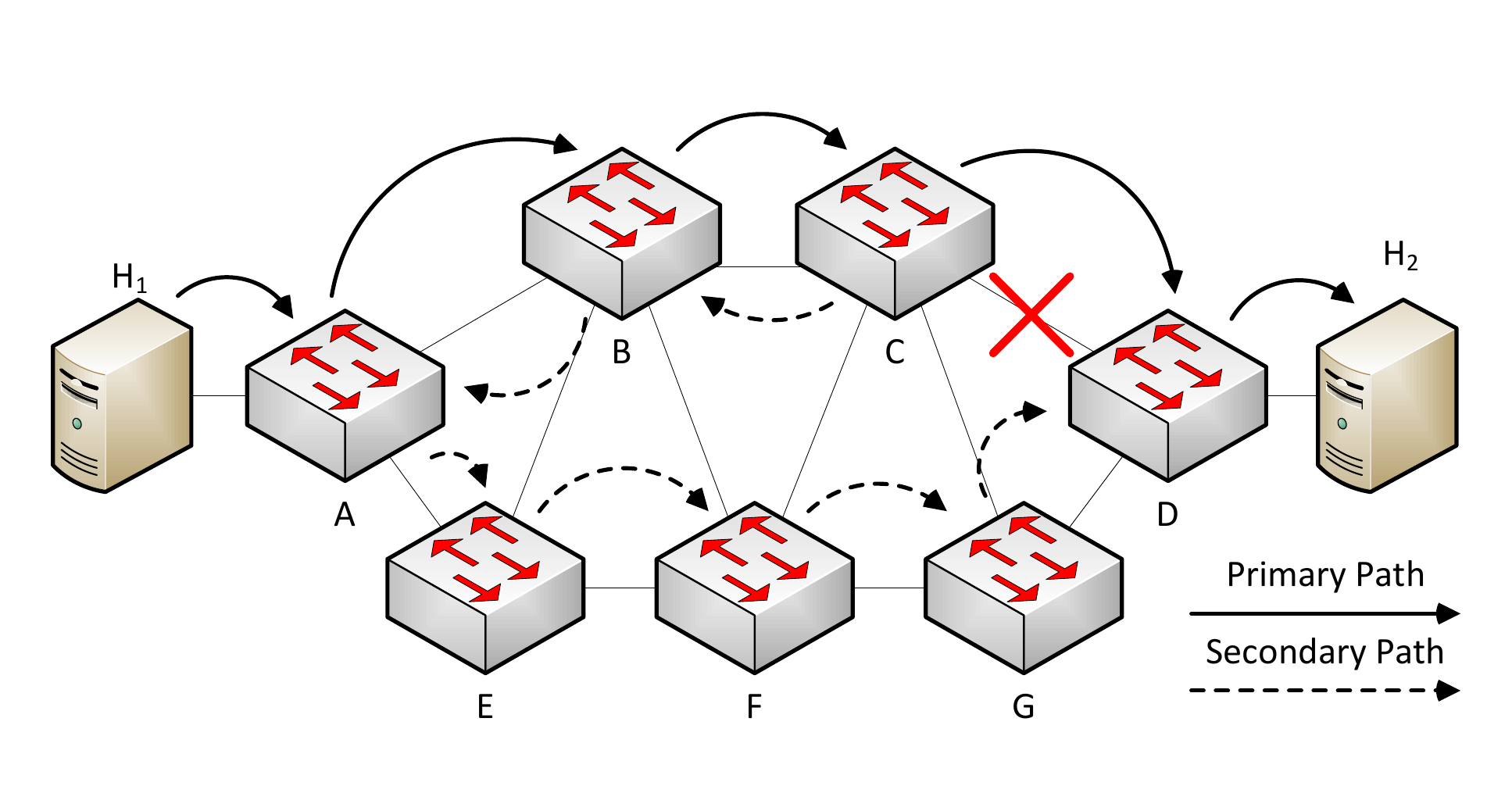}

\label{fig:DisjointPathCrankback}}

\caption{Disjoint paths and labels used in forwarding}

\label{fig:DisjointPath}
\end{figure*}

In order to evaluate failure circumventing methods as described in
the previous two sections, we have implemented an open-source prototype
OpenFlow controller module that configures a Software-Defined Network
with such backup rules. 

We have used the Ryu controller framework \cite{Ryu} as basis for
our implementation, which loads and executes our network application.
The network topology is discovered by Ryu's built-in \emph{switches
}component, while host detection occurs by a simple MAC learning procedure
in our application. 

Our application is OpenFlow 1.1+ \cite{OpenFlow11} compatible, since
it depends on the Fast-Failover Group Tables to perform the switchover
to backup paths. Tests have been performed using OpenFlow 1.3 \cite{OpenFlow13}
which is considered the current stable version of OpenFlow. 

We have used the NetworkX package \cite{NetworkX} to perform graph
creation from the learned network topology and also used it to perform
further graph manipulations and computations. We have extended the
NetworkX package in the following ways:
\begin{itemize}
\item Cleaned up the shortest path algorithms
\item Extended and standardized the (Queued) implementation of the Bellman-Ford
shortest-paths algorithm 
\item Implemented Bhandari's disjoint-paths algorithm 
\item Implemented our failure-disjoint approach
\end{itemize}
The application configures the network according to our protection
scheme, enabling it to circumvent link or node failures independent
of (slow) controller intervention. After the controller is notified
of an occurred failure, it reapplies the protection scheme to the
new network topology, reestablishing protection from future topology
failure where possible. Reconfiguration occurs without traffic interruption
using a Flow entry update strategy as explained in \cite{reitblatt2011consistent}.
Our additions to NetworkX are contributed to its source code repository.
Our open-source OpenFlow controller is published on our GitHub webpage
\cite{GitHub-OpenFlowBackupRules}.

\section{Related Work}

\label{sec:RelatedWork}

Disjoint paths, as depicted in figure \ref{fig:DisjointPathError},
are often used to preprogram alternative paths for when the primary
path of a network connections breaks. A simple and intuitive approach
for finding such disjoint paths is by using Dijkstra's algorithm \cite{Dijkstra59}
iteratively \cite{Dunn94}. At each iteration, all of the links constituting
the earlier $\{x\}_{1\leq x<k}$ disjoint paths are removed from the
network (temporarily) before Dijkstra's algorithm \cite{Dijkstra59}
is used for finding the $k$-th disjoint path. However, this iterative
approach is but a heuristic and thus cannot always return the optimal
solution even when it exists (e.g., in the presence of trap topologies
\cite{Dunn94}).

Suurballe \cite{suurballe1974disjoint} proposes an iterative scheme
for finding $k$ one-to-one disjoint paths. At each iteration, the
network is (temporarily) transformed into an equivalent network such
that the network has non-negative link weights and zero-weight links
on the links of the shortest paths tree rooted at the source node.
Dijkstra's algorithm can then be applied for finding the $k$-th disjoint
path from the knowledge of the earlier $\{x\}_{1\leq x<k}$ disjoint
paths. Bhandari \cite{Bhandari1997433} later proposed a simplification
of Suurballe's algorithm by an iterative scheme for finding the $k$-th
one-to-one disjoint path from the optimal solution of the $\{x\}_{1\leq x<k}$
disjoint paths. At each iteration, the direction and algebraic sign
of the link weight is reversed for each link of the $\{x\}_{1\leq x<k}$
disjoint paths. The network can thus contain negative link weights.
A modified Dijkstra's algorithm \cite{Bhandari1997433} or the Bellman-Ford
algorithm \cite{Bellman58,Ford62}, both usable in networks with negative
link weights, can then be applied for finding the $k$-th disjoint
path.

Both Suurballe's algorithm and Bhandari's algorithm need to be repeated
$|N|(|N|-1)$ times for finding $k$ disjoint paths between each possible
node pair, since both algorithms return only the one-to-one directed
min-sum disjoint paths between two given nodes. The Suurballe-Tarjan
algorithm \cite{Suurballe84} has reduced worst-case time complexity
for finding $k=2$ disjoint paths from one source to all possible
node pairs. The Suurballe-Tarjan algorithm also uses the equivalent
network transformation of the Suurballe algorithm to ensure that the
network contains no negative link weights in each run of the Dijkstra's
algorithm.

One of the disadvantages of using disjoint-paths based protection
is that the traffic needs to be transmitted again from the source
node using the backup path whenever the primary path fails. For example,
figure \ref{fig:DisjointPathError} shows that even when node $C$
detects the failure of link $(C,D)$, node $C$ has no means of rerouting
the packets intended for node $H_{2}$ as it is not aware of the backup
path. The only way to resolve this matter is to rely on crankback
routing as depicted in figure \ref{fig:DisjointPathCrankback}. Crankback
routing, as may be evident from the picture, implies a high network
overhead. On the other hand, our proposed algorithms enable traffic
to be rerouted directly at the current node whenever its adjacent
link or node fails, thus saving time and network resources.

There are also algorithms that propose protection schemes based on
(un)directed disjoint trees, e.g., the Roskind-Tarjan algorithm \cite{Roskind1985701}
or the Medard-Finn-Barry-Gallager algorithm \cite{Medard1999641}.
The Roskind-Tarjan algorithm finds $k$ all-to-all undirected min-sum
disjoint trees, while the Medard-Finn-Barry-Gallager algorithm finds
a pair of one-to-all directed min-sum disjoint trees that can share
links in the reverse direction. Contrary to our work, their resulting
end-to-end paths can often be unnecessarily long, which may lead to
higher failure probabilities and higher network operation costs. A
more extensive overview of disjoint paths algorithms is presented
in \cite{Kuipers12}.

In terms of work related to Software-Defined Networks, Capone et al.
\cite{capone2015detour} derive and compute an MILP formulation for
preplanning recovery paths including QoS metrics. Their approach relies
heavily on crankback routing, which results in long backup paths and
redundant usage of links compared to our approach. Their follow-up
work SPIDER \cite{cascone2015spider} implements the respective failure
rerouting mechanism using MPLS tags. The system relies heavily on
OpenState \cite{bianchi2014openstate} to perform customized failure
detection and data plane switching, making it incompatible with existing
networks and available hardware switches. Furthermore, the system
does not distinguish between link and node failures as our approach
does.

IBSDN \cite{ibsdn} achieves robustness through running a centralized
controller in parallel with a distributed routing protocol. Initially,
all traffic is forwarded according to the controller's configuration.
Switches revert to the path determined by the traditional routing
protocol once a link is detected to be down. The authors omit crankback
paths through crankback detection using a custom local monitoring
agent. The proposed system is both elegant and simple, though does
require customized hardware, since switches need to connect to a central
controller, run a routing protocol, and implement a local agent to
perform crankback detection. Moreover, the time it takes the routing
protocol to converge to the post-failure situation may be long and
cannot outpace a preconfigured backup plan.

Braun et al. \cite{LFA-SDN} apply the concept of Loop-Free Alternates
(LFA) from IP networks to SDNs, where nodes are preprogrammed with
single-link backup rules when not creating loops. Through applying
an alternative loop-detection method more backup paths are found than
using traditional LFA, although full protection requires topological
adaptations.

\section{Conclusion}

\label{sec:Conclusion}

In this paper we have derived, implemented and evaluated algorithms
for computing an all-to-all network forwarding configuration capable
of circumventing link and node failures. Our algorithms compute forwarding
rules that include failure-disjoint backup paths offering preprogrammed
protection from future topology failures. Through packet labelling
we guarantee correct and loop-free detour forwarding. The labeling
technique allows packets to return on primary paths unaffected by
the failure and carries information used to upgrade link-failures
to node-failures when applicable. Furthermore, we have implemented
a proof-of-concept network controller that configures OpenFlow-based
SDN switches according to this approach, showing that these types
of failover techniques can be applied to production networks. 

Compared to traditional link- or node-disjoint paths, our method shows
to have significantly shorter primary and backup paths. Furthermore,
we observe significantly less crankback routing when backup paths
are activated in our approach. Besides shorter paths, our approach
outperforms traditional disjoint path computations in terms of respectively
the needed Flow and Group table configuration entries by factors up
to $20$ and $1.9$. Our approach allows packets that encounter a
broken link or node along their path, to travel the second-to-shortest
path to their destination taken from the node where the link or node
failure is detected. We apply Software-Defined Networking, specifically
the OpenFlow protocol, to configure computer networks according to
the derived protection scheme, allowing them to continue functioning
without (slow) controller intervention. After the network controller
is notified of the link or node failure it reconfigures the network
by applying the protection scheme to its newly established topology,
therewith reassuring protection from future topology failure within
reasonable time.

\bibliographystyle{IEEEtran}
\bibliography{References,rfc}

\end{document}